\documentclass[manuscript=article]{achemso}
\usepackage[version=3]{mhchem}
\usepackage{subcaption}
\usepackage{stfloats} 
\usepackage{graphicx}                     
\usepackage{booktabs}    
\usepackage{array}
\usepackage{tabularx}
\newenvironment{DIFnomarkup}{}{}
\newcommand*\ruo{\ce{Rb2Ru2O7\!.\!H2O}}

\author{Krystof Chrappova}
\affiliation{
School of Chemistry, University of Bristol, Cantock’s Close, Bristol BS8 1TS, UK}
\author{Jeremiah P. Tidey}
\affiliation{Department of Physics, University of Warwick, Coventry, CV4 7AL, UK}
\author{Christopher Bell}
\affiliation{School of Physics, University of Bristol, Tyndall Avenue, Bristol BS8 1TL, UK}
\author{Simon R. Hall}
\affiliation{
School of Chemistry, University of Bristol, Cantock’s Close, Bristol BS8 1TS, UK}
\email{simon.hall@bristol.ac.uk}

\title[Structure and Magnetism of \ruo]{Magnetism and 3D Electron Diffraction Solution of Hydrated Rubidium–Ruthenium Oxide \ruo}

\abbreviations{3D ED}

\begin{document}

\begin{abstract}
The crystal structure of \ruo\ was determined by three-dimensional electron diffraction from the individual crystallites of a solid-state powder product. \ruo\ crystallizes in space group \textit{C}2/\textit{c} ($a=7.841(3)$ \AA, $b=12.500(3)$ \AA, $c=8.392(2)$ \AA, $\beta=93.57(4)^\circ$, Z=4). The structure contains infinite chains that run normal to the (101) plane and consist of alternating \ce{RuO6} octahedra and square-pyramidal \ce{RuO5} units connected via shared O–O edges. Magnetic properties were measured on the bulk powder, showing a diamagnetic baseline from 300 to 60~K with a small Curie tail below 55~K. The magnetic moment, calculated from the 1.8~K isotherm, saturates at $M=4.4\times10^{-3}\,\mu_\mathrm{B}\,\mathrm{Ru}^{-1}$, much less than would be expected for $S=1$ ruthenium.  Bond-valence-sum analysis indicates high-valent Ru, and the near-diamagnetic response is consistent with the edge-sharing Ru–Ru motif, where weak direct Ru–Ru overlap yields a local singlet.

\end{abstract}

Ruthenium oxides display a wide range of unusual electronic ground states.  
Among alkaline metal ruthenates, $\mathrm{SrRuO_3}$ is an itinerant ferromagnet ($T_\mathrm{Curie}=150\,$ K), whereas its iso-structural analogue $\mathrm{CaRuO_3}$ remains paramagnetic when undoped.\cite{Jeong_2013,Shepard_1996} $\mathrm{Sr_2RuO_4}$ exhibits unconventional superconductivity (superconducting $T_\mathrm{c}=1.5\,$ K) \cite{Maeno_1994}.

In the alkali-ruthenate family, intermediate valence $\mathrm{Ru^{4+/5+}}$ allow diverse extended frameworks. Two-dimensional honeycomb layers have been reported for $\mathrm{Li_2RuO_3}$ and $\mathrm{Na_2RuO_3}$, and isolated octahedral chains are present in $\mathrm{Li_3RuO_4}$ \cite{Miura_2007,Wang_2014,Alexander_2003}.  

Here we present the previously unknown hydrated oxide \ruo, whose structure comprises isolated zig‑zag chains of alternating edge‑sharing \ce{RuO6} octahedra and \ce{RuO5} square pyramids. The compound was solved by three‑dimensional electron diffraction (3D ED) and its magnetic properties were determined on the bulk polycrystalline solid-state product.
\begin{figure}[!t]
  \centering
  \begin{subfigure}{0.6\linewidth}
    \centering
    \includegraphics[width=0.9\linewidth]{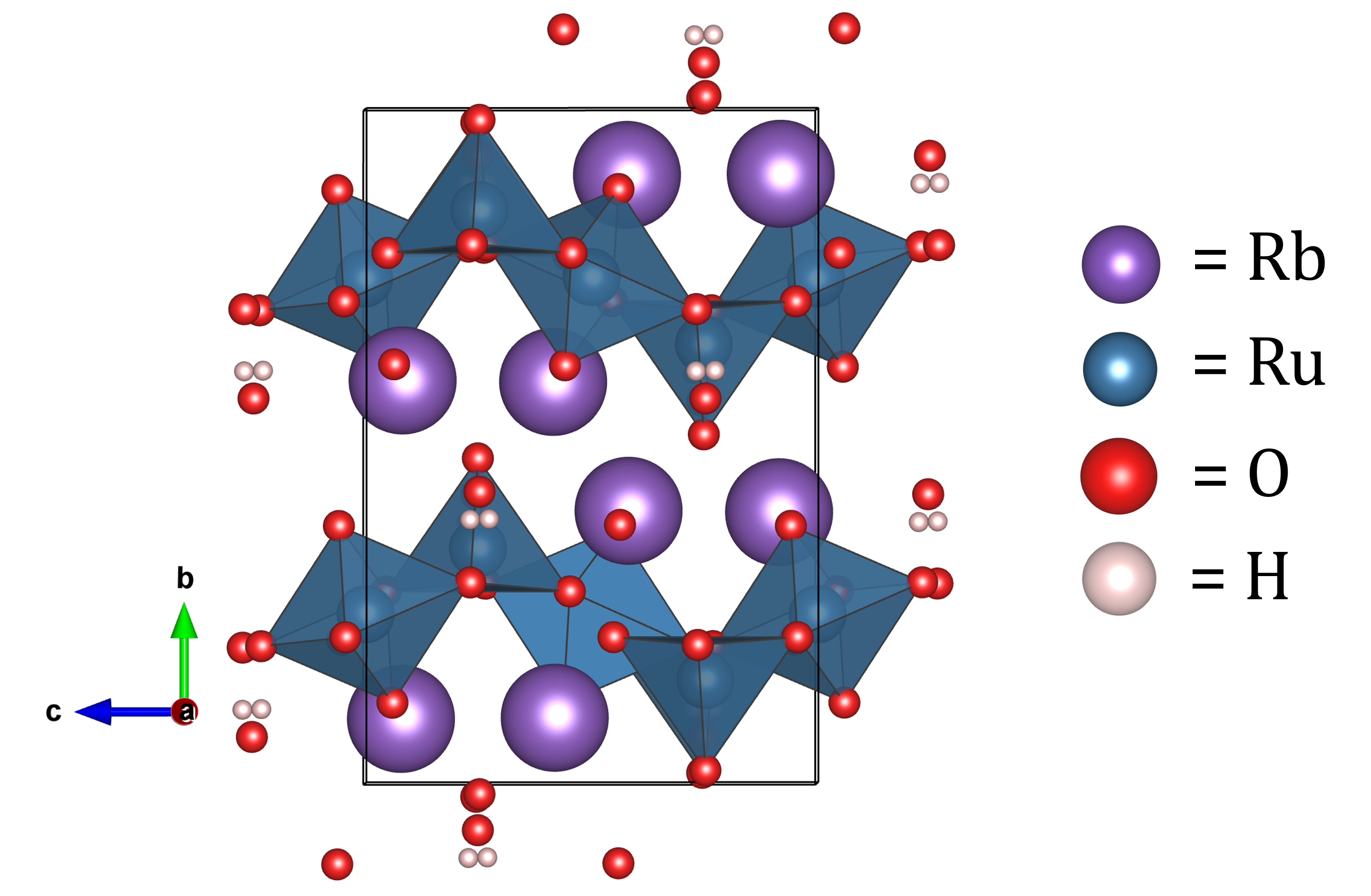}
    \caption{}
  \end{subfigure}

  \begin{subfigure}{0.6\linewidth}
    \centering
    \includegraphics[width=0.5\linewidth]{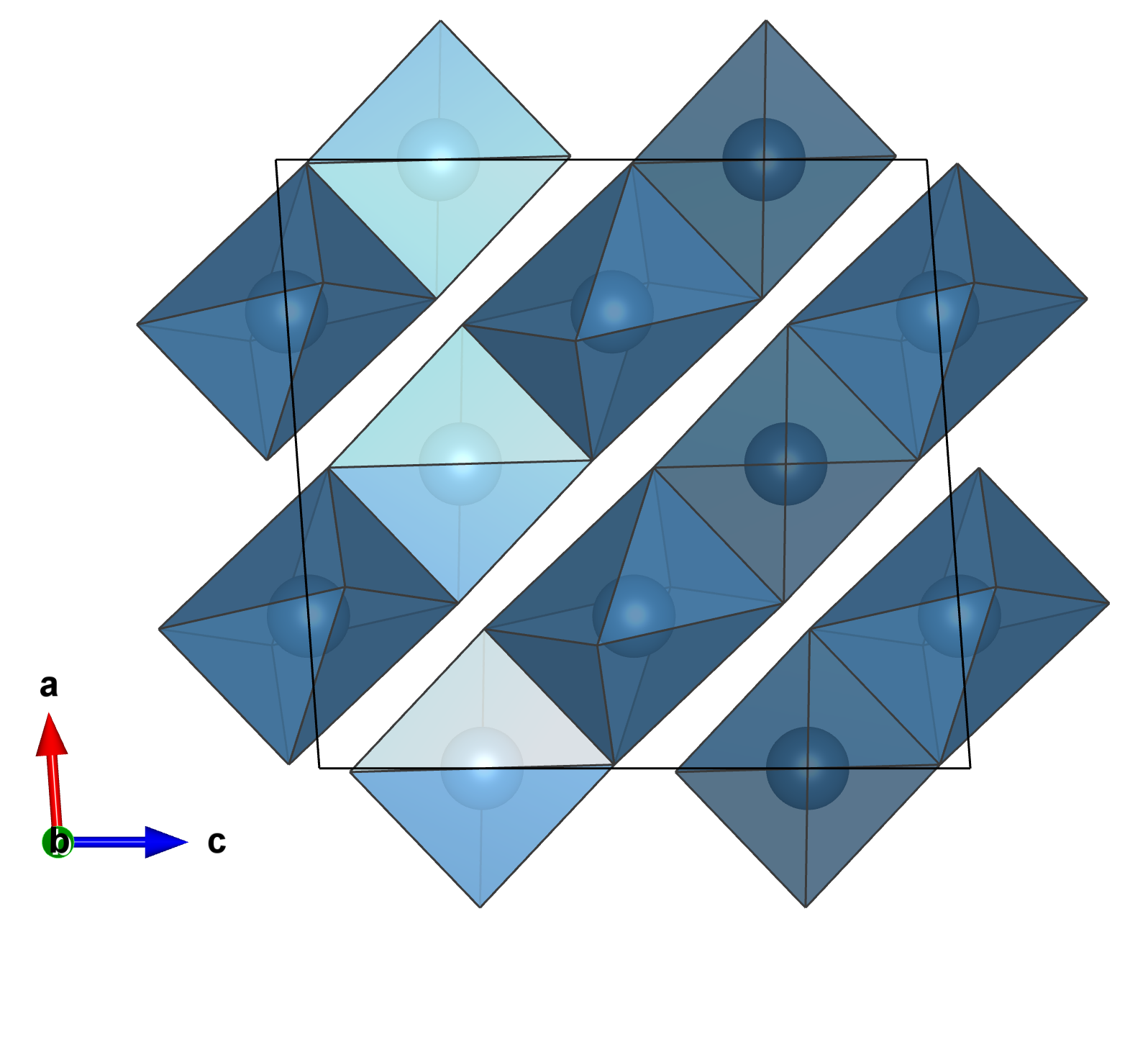}
    \caption{}
  \end{subfigure}

  \begin{subfigure}{0.6\linewidth}
    \centering
    \includegraphics[width=0.6\linewidth]{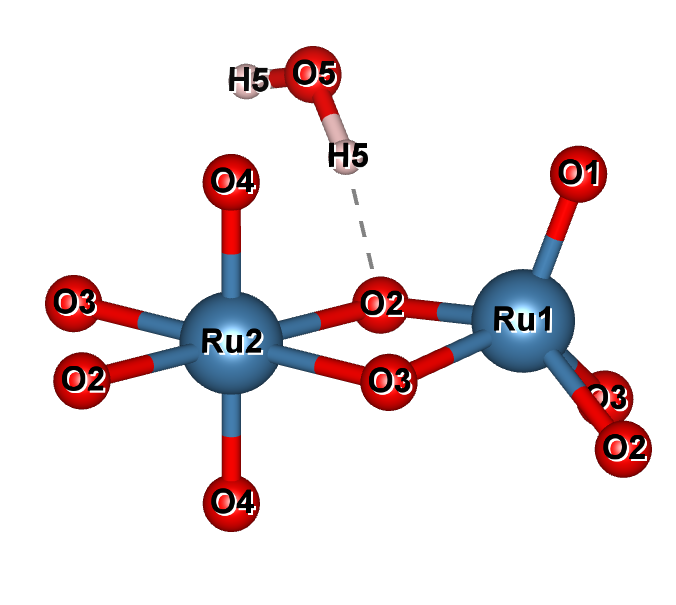}
    \caption{}
  \end{subfigure}

  \caption{Polyhedral representation of (a) 3D ED structure solution projected onto $(100)$ for \ruo\ in the monoclinic \textit{C}2/\textit{c} space group ($a=7.841(3)$ \AA, $b=12.500(3)$ \AA, $c=8.392(2)$, Z = 4, $\beta=93.57(4)^\circ$), (b) Zig-zag chain of alternating \ce{RuO6}/\ce{RuO5} projected onto $(010)$. (c) Edge-sharing unit that comprises the nominal molecule labeled by crystallographically unique atoms.}
  \label{fig:stacked}
\end{figure}

\begin{figure}[!t]
  \centering
  \includegraphics[width=0.8\linewidth]{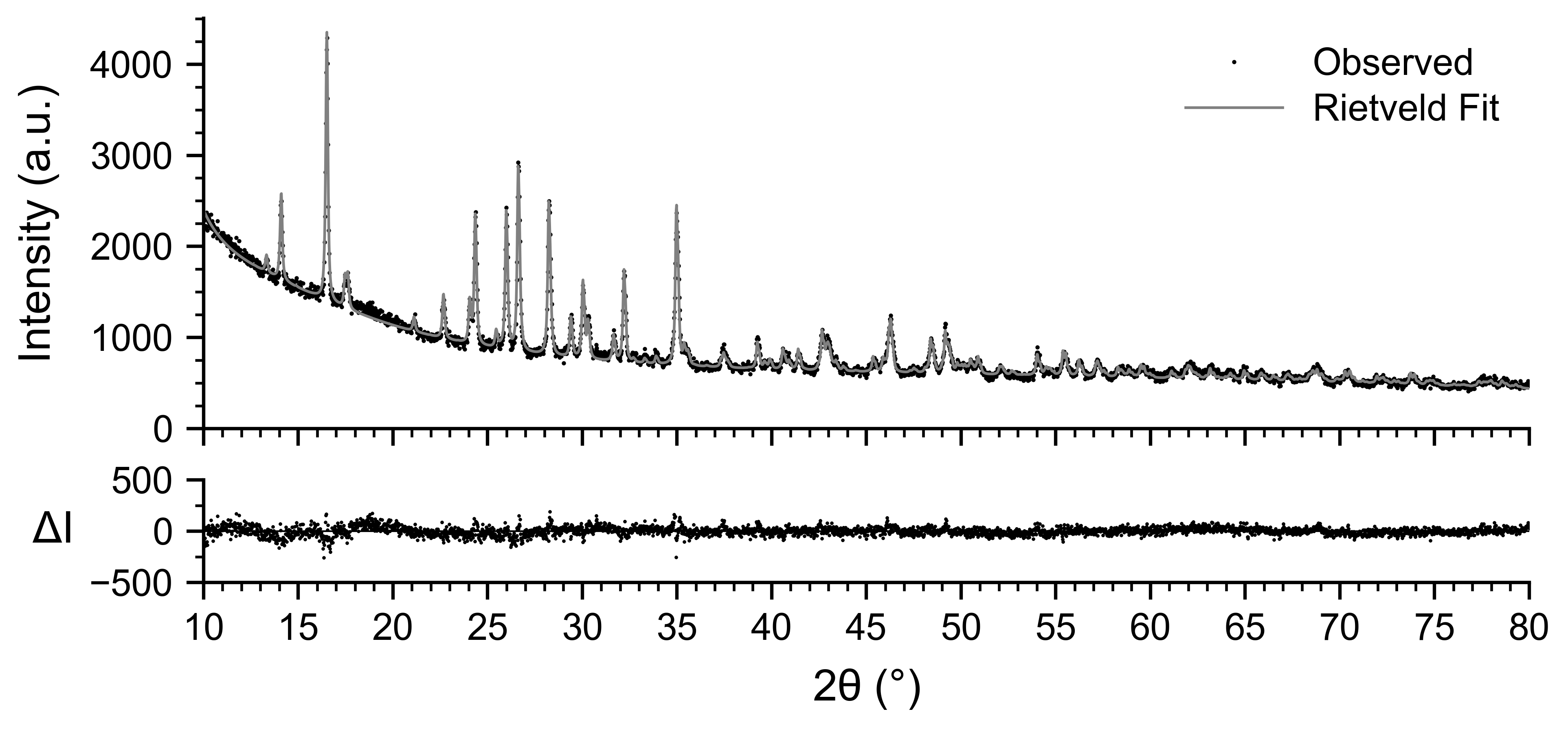}
  \caption{PXRD pattern of bulk \ruo\ with a Rietveld refinement using the structure solved by 3D ED (ICSD deposition: CSD 2514706) without internal standard. Rietveld refinement: $\chi^2 = 1.8$, $R_\mathrm{wp} = 4.5$, $R_\mathrm{exp} = 3.4$.}
  \label{fig:PXRD}
\end{figure}

A polycrystalline sample of \ruo\ was obtained by reacting ground \ce{Rb2CO3} (1.5\, mmol) with \ce{RuO2} (0.75\, mmol) in an alumina crucible at $1000\,^\circ\mathrm{C}$ (4\,h; $5\,^\circ\mathrm{C}\,\mathrm{min}^{-1}$ ramp). The structure (Figure \ref{fig:stacked}) was solved from micron-sized crystals (S1) using continuous rotation, selected area 3D ED performed at 100(5) K. Rietveld refinement using the 3D ED model fits the laboratory powder X-ray diffraction (PXRD) pattern of the bulk sample well (Figure \ref{fig:PXRD}). Scanning electron microscopy with energy-dispersive X-ray spectroscopy  shows co-localisation of both metals used in the synthesis (S3-S4). Transmission electron microscopy reveals the powder consists of polycrystalline, needle-like particles up to $6\,\mu\mathrm{m}$ in length, and the corresponding selected area electron diffraction shows rings that can be indexed to the lattice planes in the 3D ED structure (S5, Table S3).

The results from 3D ED show two crystallographically independent Ru sites are present in the crystal structure (Figure~\ref{fig:stacked} (c)). Ru1 is five-coordinate, forming a distorted square pyramid with an apical Ru1–O1 bond of 1.656(8)\,\AA\ and basal bonds Ru1–O3 = 1.896(6)\,\AA\ ($\times 2$) and Ru1–O2 = 1.910(6)\,\AA\ ($\times 2$). Ru2 is six-coordinate, with two short axial contacts Ru2–O4 = 1.741(6)\,\AA\ and four equatorial contacts Ru2–O3 = 1.994(6)\,\AA\ ($\times 2$) and Ru2–O2 = 2.032(5)\,\AA\ ($\times 2$). The polyhedra share the O2–O3 edge, giving Ru1\,$\cdots$\,Ru2 = 3.0454(14)\,\AA. Edge-sharing continues normal to the $(101)$ plane (Figure~\ref{fig:stacked}).

\begin{figure}[!t]
  \centering

  \begin{subfigure}{0.8\linewidth}
    \centering
    \includegraphics[width=\linewidth]{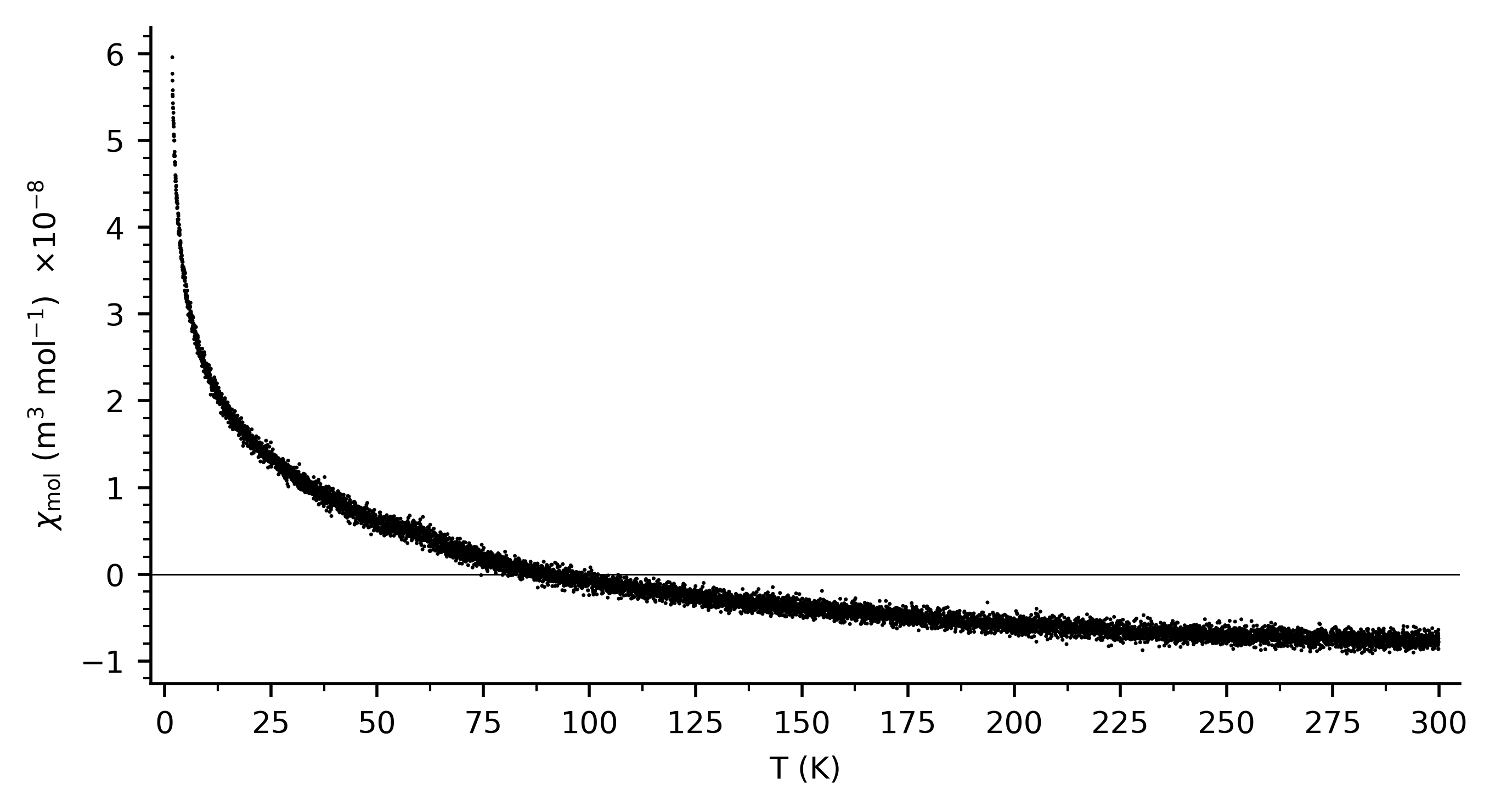}
    \caption{}
  \end{subfigure}

  \vspace{6pt}

  \begin{subfigure}{0.8\linewidth}
    \centering
    \includegraphics[width=\linewidth]{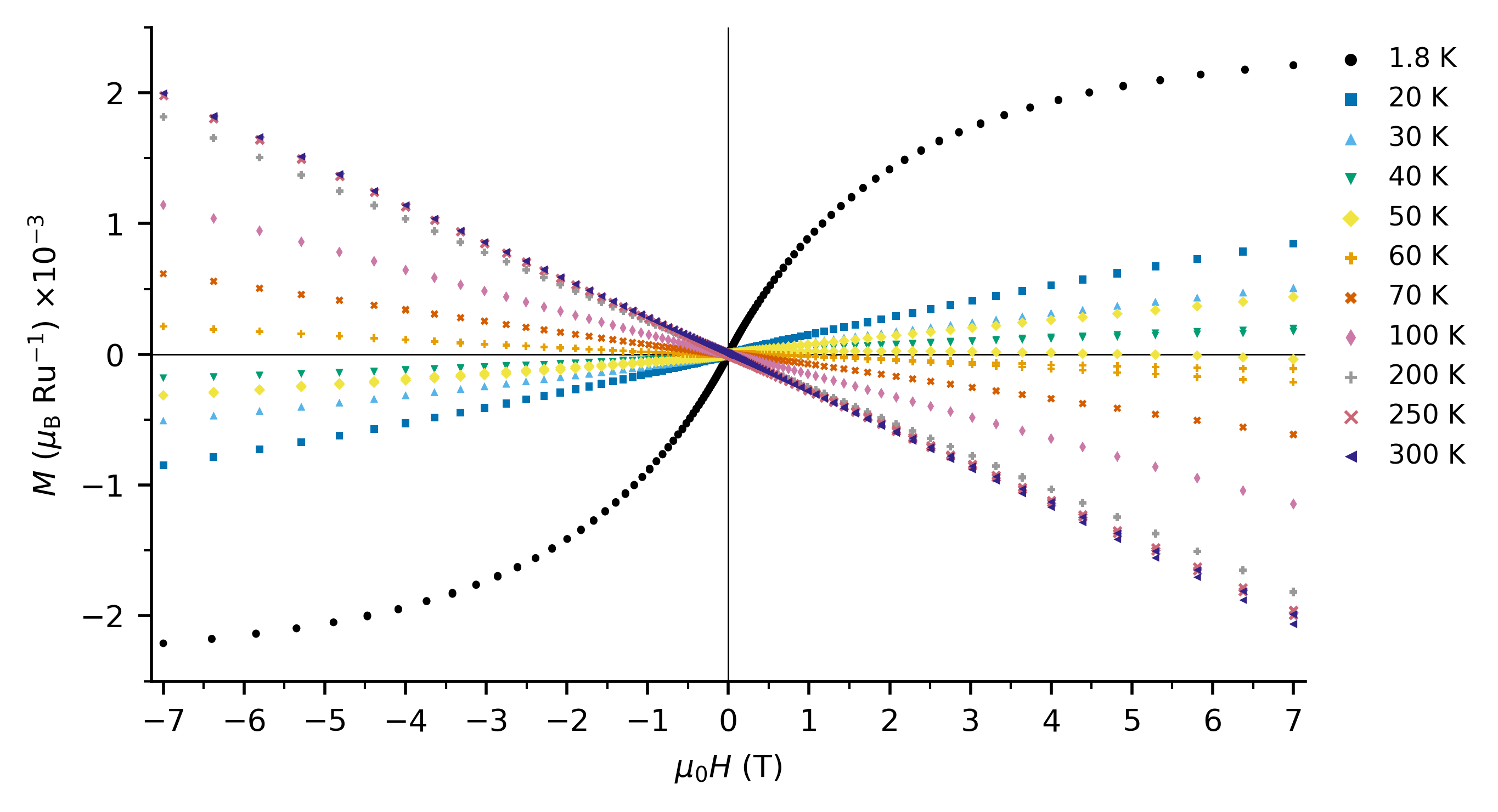}
    \caption{}
  \end{subfigure}

  \caption{(a) Temperature dependence of the molar susceptibility, $\chi(T)$, for bulk polycrystalline \ruo, zero field cooled with applied field of 1 mT. (b) Isothermal magnetization, $M(\mu_{0}H)$, at selected temperatures. All $\chi$ and $M$ data are normalized to the crystalline \ruo\ mass fraction obtained by Rietveld refinement with an \ce{Al2O3} internal standard ($f_\mathrm{cryst}=0.25$) and values are reported per Ru in the crystalline phase.}
  \label{fig:squid_data}
\end{figure}

Figure~\ref{fig:squid_data} (a) shows the temperature dependence of the molar susceptibility. A temperature-independent baseline $\chi_{0}=-1.06\times10^{-8}\,\mathrm{m^{3}\,mol^{-1}}$ persists from 300 to 60~K, followed by a Curie upturn below $\sim55$~K. A linearization of $[\chi(T)-\chi_{0}]^{-1}$ against temperature over 5–55~K yields a Curie constant $C=1.32\times10^{-6}\,\mathrm{m^{3}\,mol^{-1}\,K}$ and an effective moment $\mu_{\mathrm{eff}}=0.65\,\mu_\mathrm{B}\,\mathrm{Ru}^{-1}$. The 1.8~K, 7~T isotherm saturates at $M=4.4\times10^{-3}\,\mu_\mathrm{B}\,\mathrm{Ru}^{-1}$, implying that only a small fraction of Ru sites carry $S=\tfrac12$ (or $S=1$) moments, or that the moments form some kind of anti-ferromagnetic-like order whose anti-parallel moments are not significantly canted at an applied field of 7 T. Rietveld refinement with an \ce{Al2O3} internal standard shows that $27~\%$ of the bulk mass is crystalline \ruo (Figure S2). All $\chi$ and $M$ data are normalized to the crystalline \ruo\ mass fraction within the sample. Thus trivial dilution by amorphous material has already been removed. Therefore, the vanishing net moment is intrinsic to \ruo.

Bond-valence-sum (BVS) analysis was carried out for both Ru coordination polyhedra. Using Brese–O'Keeffe parameters for Ru$^{4+}$ ($R_{0}=1.834$\,\AA, $B=0.37$\,\AA) gives $V_\text{Ru1}=4.94$ and $V_\text{Ru2}=5.04$\,v.u.\,\cite{Brese_1991} Parameters for higher oxidation states are not available using this reference.
Using the Gagné–Hawthorne Ru$^{4+}$ parameters ($R_{0}=1.833$\,\AA, $B=0.366$\,\AA) produces similar results $V_\text{Ru1}=4.93$ and $V_\text{Ru2}=5.02$ v.u., whereas their Ru$^{5+}$ set ($R_{0}=1.894$\,\AA, $B=0.346$\,\AA) increases the sum to $V_\text{Ru1}=5.89$ and $V_\text{Ru2}=5.95$ v.u.\,\cite{Gagn__2015} Parameters for Ru$^{6+}$ are not available. Because the Gagné–Hawthorne parameter sets were calibrated predominantly for regular RuO$_6$ octahedra, the distortion of the Ru2 octahedron and the short apical bond in Ru1 inflate the valence sum and the BVS is in this case semi-quantitative allowing for both +5 and +6 formal charge.\,\cite{Gagn__2020}
\par

Both BVS-permitted oxidation states can plausibly produce Ru-Ru bonding combinations that yield a local singlet consistent with the observed near-diamagnetism. For Ru$^{5+}$ (d$^{3}$ + d$^{3}$), edge-sharing Ru pairs permit three $t_{2g}$–$t_{2g}$ overlaps that form bonding combinations denoted $\sigma,\pi,\delta$, as has been previously shown for the $\mathrm{Li_2RuO_3}$ system.\cite{Miura_2009} Six electrons can then fill this bonding set as $\sigma^{2}\pi^{2}\delta^{2}$, giving a closed-shell $S=0$.
For the Ru$^{6+}$ (d$^{2}$ + d$^{2}$) analogue, either a molecular orbital configuration that results in S=1 ($\sigma^{2}\pi^{1}\delta^{1}$) or closed shell configuration ($\sigma^{2}\pi^{2}\delta^{0}$) is possible. However, only the latter is consistent with the observed magnetism.
The Ru1–Ru2 separation of 3.0298(16)\,\AA\ lies within the range where weak direct overlap has been invoked (viz., 2.76–3.06\,\AA\ in $\mathrm{Y_5Ru_2O_{12}}$).\cite{Sanjeewa_2020,Reeves_2019}

We conclude, therefore, that the magnetism here is likely governed through overlap of Ru–Ru neighbors. The open-shell Ru$^{6+}$ molecular orbital configuration or Ru$^{4+}$/Ru$^{6+}$ disproportionation would contradict either the Curie constant or the single set of Ru–O distances and are thus ruled out.

\begin{acknowledgement}
The authors thank the NCS for granting rapid access to the National Electron Diffraction Facility under EPSRC funding (EP/X014606/1 \& EP/X014444/1, A National Electron Diffraction Facility for Nanomaterial Structural Studies), as well as the University of Warwick X-ray Research Technology Platform for provision of further analysis facilities. The Gatan Elsa specimen holder was procured under EPSRC funding (EP/R019428/1). JPT also thanks Dr.rer.nat. Lukáš Palatinus for his continued expert advice and fruitful discussions. JPT thanks Dr Struan Simpson for fruitful discussion and Prof. Jeremy Sloan for his generous provision of the Gatan Elsa.
This work was supported by the Henry Royce Institute for advanced materials through the Equipment Access Scheme enabling access to the Magnetic Property Measurement System at Cambridge; Cambridge Royce facilities grant EP/P024947/1 and Sir Henry Royce Institute - recurrent grant EP/R00661X/1. KC thanks the EPSRC TECS CDT EP/S024107/1 for a PhD scholarship. The authors thank Jan Maurycy Uszko for running TEM analysis, and Dr Cheng Liu for running the  magnetic measurements, Dr. Chris Russell for useful discussions, and Dr. Ella Gale for supervisory support. 
\end{acknowledgement}

\begin{suppinfo}
Experimental procedures, 3D ED details for \ruo ~(ICSD deposition: CSD  2514706), SEM, TEM, and PXRD with internal standard data.
\end{suppinfo}


{\large \bf Supplementary Information: Magnetism and 3D Electron Diffraction Solution of Hydrated Rubidium–Ruthenium Oxide \ruo }

This PDF contains synthetic procedure, crystallographic, magnetic property measurements, and electron microscopy details.

\section{Synthetic Procedure}
Rubidium carbonate (0.347g, 1.5 mmol) and ruthenium oxide (IV) (0.100g, 0.75 mmol) were ground together with a mortar and pestle until a homogeneous powder with uniform color was obtained. The boiling point of rubidium carbonate is lower than the temperature at which the product is synthesized. For this reason, an excess of rubidium carbonate was used in the synthesis. Synthesis was carried out in an alumina crucible in a box furnace at 1000$^\circ$C,  for 4 hours at ramp rate 5$^\circ$C min$^{-1}$. The product was as a dark brown powder obtained after air-cooling in the furnace.

\section{Crystallographic Details}
For the 3D ED experiment, the sample was dispersed dry as received onto a copper-supported holey amorphous carbon TEM grid and loaded at 100 K via a high-tilt Gatan Elsa specimen holder into a Rigaku XtaLAB Synergy-ED electron diffractometer, operated at 200 kV and equipped with a Rigaku HyPix-ED hybrid pixel array area detector. Data were collected on various crystallites as single-rotation scans collecting 0.25$^\circ$ frames using CrysAlisPRO system (CCD 1.171.43.129a 64-bit (release 29-06-2024))\cite{3DED_1} using continuous rotation electron diffraction with a selected area aperture of 2 $\mu$m apparent diameter. Further experimental details are provided in Table \ref{tab:S1}. 
All single crystal component datasets were individually indexed and integrated, prior to the merging of three suitable datasets for scaling, using CrysAlisPRO (version 1.171.44.113a)\cite{3DED_1} ; no absorption corrections or outlier rejections were applied. The structure was solved using ShelXT\cite{Sheldrick_2015} and initially refined using Olex2.refine in the kinematic approximation, applying an extinction correction to broadly account for multiple scattering as implemented in Olex2 (version 1.5-ac7-016, compiled 2025.05.29 svn.r8cd99b3d for Rigaku Oxford Diffraction, GUI svn.r7254)\cite{Bourhis_2015, Dolomanov_2009} and using published scattering factors\cite{Saha_2022}. 
The resulting atomic coordinates  were used as the starting point for a full dynamical refinement against the two highest quality data collections which were individually reduced and scaled without outlier rejection in CrysAlisPRO (version 1.171.44.128a)\cite{3DED_1}. The dynamical reflection files were concurrently imported into JANA2020 (version 2.1)\cite{Pet_ek_2023}, taking as reference unit cell parameters those for the highest quality component, and refinement was performed in the presence of ribbon models of thickness and anisotropic models of incoherent mosaicity for both datasets. All non-H atoms are refined anisotropically and the proton, earlier located in the residual difference map, was refined isotropically, all without the application of any restraints.
We take as our final model to be the dynamically refined model, yet deposit CIFs for both outputs in the ICSD: deposition numbers CSD 2464145 and 2514706 for the SHELX-compatible kinematical (Table \ref{tab:S1}) and JANA2020 dynamical ( Table \ref{tab:S2}) refinements, respectively. These contain complete experimental and refinement information along with appropriate structure factors and, in the former case, an embedded .RES file.

\begin{figure}
  \centering
  \begin{subfigure}{\linewidth}
    \centering
    \includegraphics[width=0.6\linewidth]{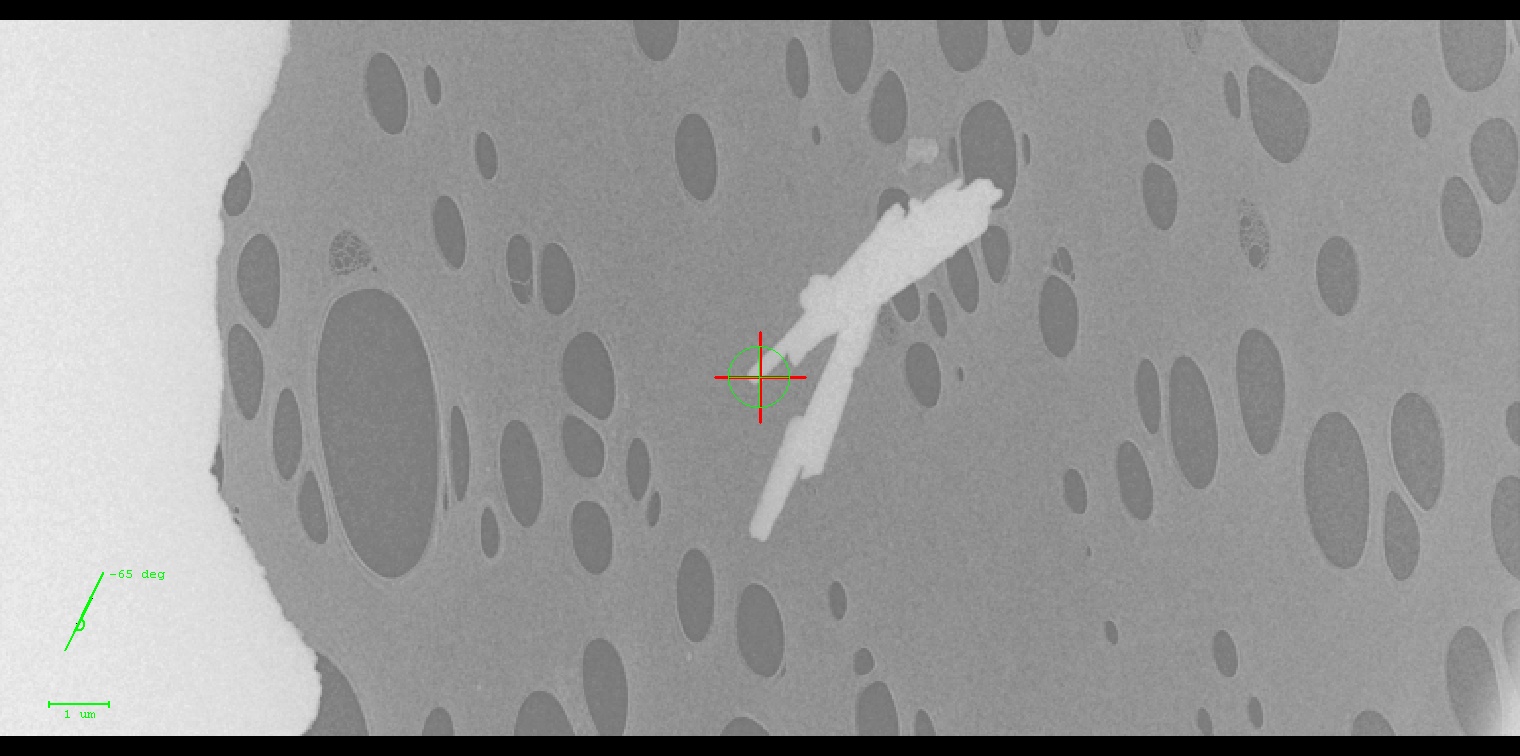}
    \caption{}
  \end{subfigure}

  \begin{subfigure}{\linewidth}
    \centering
    \includegraphics[width=0.6\linewidth]{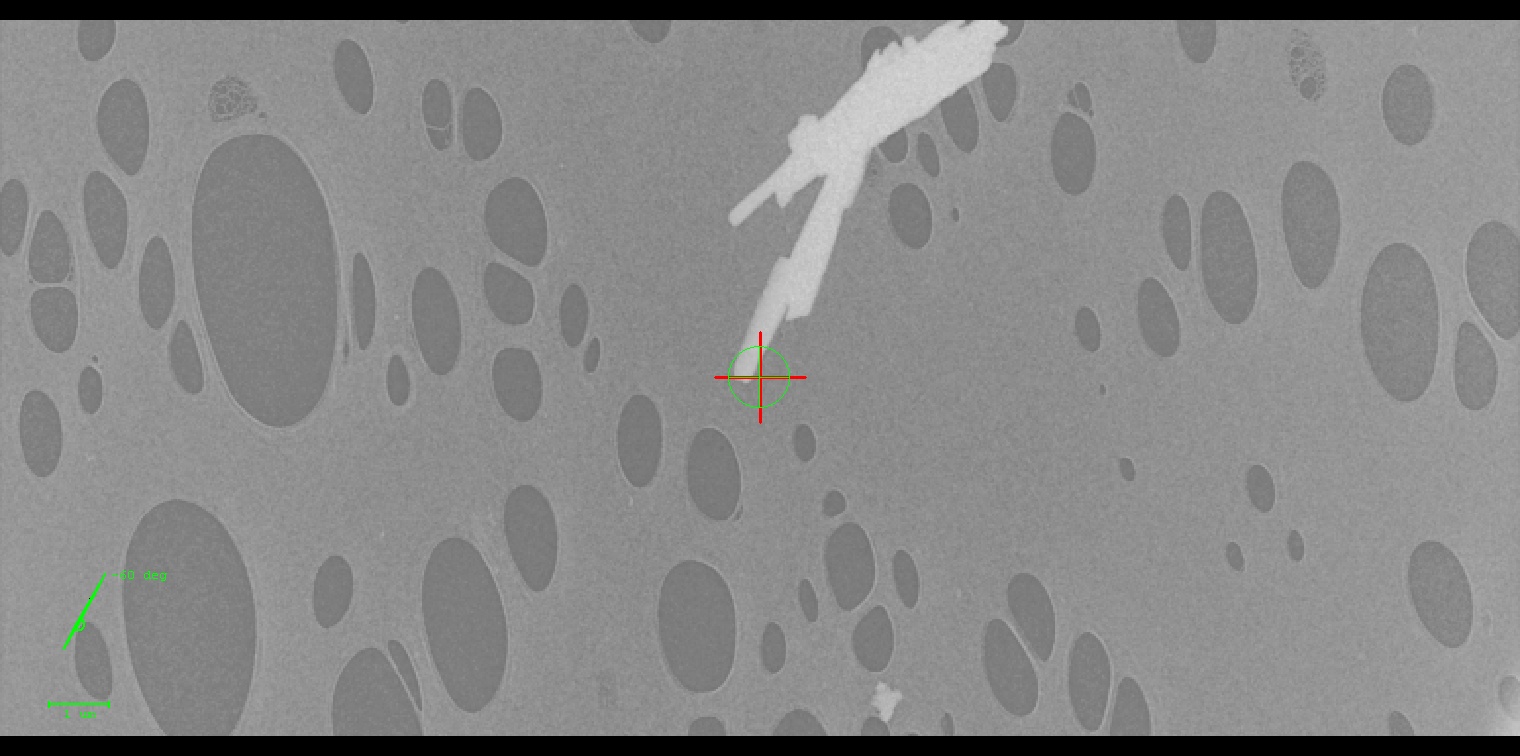}
    \caption{}
  \end{subfigure}

  \begin{subfigure}{\linewidth}
    \centering
    \includegraphics[width=0.6\linewidth]{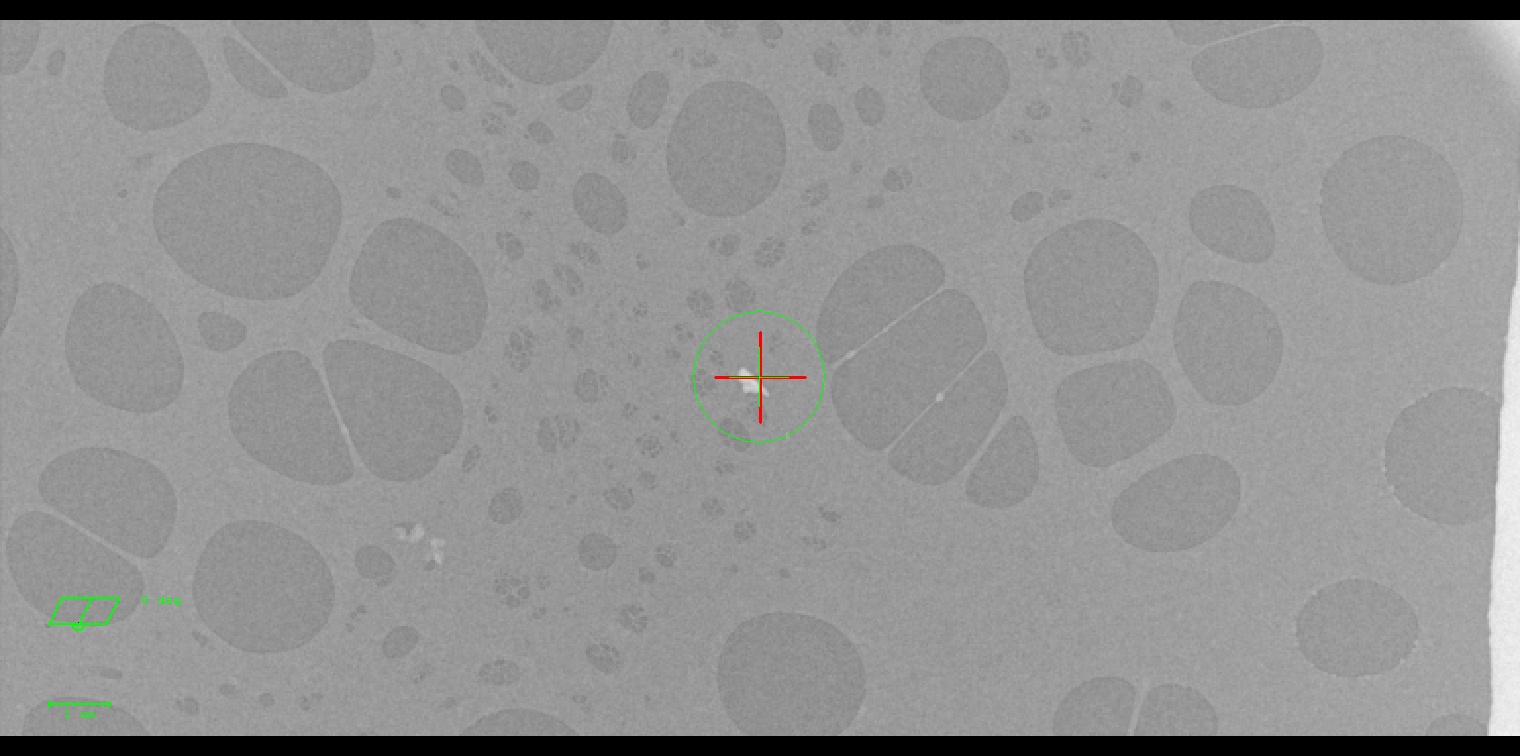}
    \caption{}
  \end{subfigure}

  \caption{Images of the three crystallites from which the data are collected. The 1 $\mu$m scale bar is shown bottom-left of each image, alongside the tilt angle at which the image is taken, while the green circle in the center represents the approximate location of the selected area aperture used in each experiment.}
  \label{figstacked}
\end{figure}

\begin{DIFnomarkup}
\begin{table}[htbp]
  \centering
  \caption{Experimental details for the kinematic refinement against the merging of datasets collected on three distinct crystallites. The according CIF is deposited in the ICSD with Deposition Number CSD 2464145.}
  \label{tab:S1}
    \begin{tabularx}{\linewidth}{@{}l >{\raggedright\arraybackslash}X@{}}
    \toprule
    \multicolumn{2}{@{}c@{}}{\textbf{Crystal data}} \\
    \midrule
    Chemical formula                       & \ce{Rb2Ru2O7(H2O)}                  \\
    $M_r$                                  & 503.8                              \\
    Crystal system, space group            & Monoclinic, \textit{C}2/\textit{c}  \\
    Temperature / K                        & 100(5)                              \\
    $a$, $b$, $c$ / \AA                      & 7.8575(18), 12.485(3), 8.3864(19)   \\
    $\alpha$, $\beta$, $\gamma$ / $^\circ$ & 90, 94.10(2), 90                    \\
    $V$ / \AA$^{3}$                           & 820.6(3)                            \\
    $Z$                                    & 4                                   \\
    Radiation type                         & Electron, $\lambda$ = 0.0251 \AA     \\
    \addlinespace
    \multicolumn{2}{c}{\textbf{Data collection}} \\
    \midrule
    Scan range / $^\circ$                  & See \texttt{\_diffrn\_measurement\_details} in CIFs for individual component angular ranges \\
    Measured, independent, observed $[I \ge 2u(I)]$ reflections
                                           & 5593, 999, 790                    \\
    $R_\mathrm{int}$                       & 0.209                               \\
    $(\sin\theta/\lambda)_{\max}$ / Å$^{-1}$ & 0.674                               \\
    \addlinespace
    \multicolumn{2}{@{}c@{}}{\textbf{Refinement}} \\
    \midrule
    $R_1$, $wR_2$ ([$F^2 > 2\sigma(F^2)$])   & 0.1455, 0.2949                      \\
    $R_1$, $wR_2$ (all)               & 0.1723, 0.3110                      \\
    GoF ($S$) (incl.,excl. restraints)       & 1.0495, 1.0562                     \\
    No.\ reflections                       & 999                                 \\
    No.\ parameters                        & 63                                  \\
    No.\ restraints                        & 12                                  \\
    $\Delta\phi_{\max}$, $\Delta\phi_{\min}$ / e\AA$^{-1}$ & 1.69, $-1.84$                       \\
    \bottomrule
  \end{tabularx}

  \vspace{0.5em}

\end{table}
\end{DIFnomarkup}

\begin{table}[htbp]
  \centering
  \caption{Experimental details for the dynamical co-refinement against distinct datasets collected on two crystallites. The according CIF is deposited in the ICSD with Deposition Number CSD 2514706.}
  \label{tab:S2}
    \begin{tabularx}{\linewidth}{@{}l >{\raggedright\arraybackslash}X@{}}
    \toprule
    \multicolumn{2}{c}{\textbf{Crystal data}} \\
    \midrule
    Chemical formula                       & \ce{Rb2Ru2O7(H2O)}                  \\
    $M_r$                                  & 503.8                              \\
    Crystal system, space group            & Monoclinic, \textit{C}2/\textit{c}  \\
    Temperature / K                        & 100(5)                              \\
    $a$, $b$, $c$ / \AA                      & 7.841(3), 12.500(3), 8.392(2)   \\
    $\alpha$, $\beta$, $\gamma$ / $^\circ$ & 90, 93.57(4), 90                    \\
    $V$ / \AA$^{3}$                           & 820.9(4)                            \\
    $Z$                                    & 4                                   \\
    Radiation type                         & Electron, $\lambda$ = 0.0251 \AA     \\
    \addlinespace
    \multicolumn{2}{c}{\textbf{Data collection}} \\
    \midrule
    Measured, observed $[I \ge 2u(I)]$ reflections (combined(component 1+2))                & 4241 (2236+2005), 2682 (1501+1181) \\
    $R_\mathrm{int}$  (component 1; 2)                     & 0.1514, 0.2044                               \\
    $(\sin\theta/\lambda)_{\max}$ integ. limit / \AA$^{-1}$ & 0.674                               \\
    \addlinespace
    \multicolumn{2}{c}{\textbf{Refinement}} \\
    \midrule
    $(\sin\theta/\lambda)_{\max}$ data cutoff / $\mathrm{\AA}^{-1}$ & 0.70 \\
    Completeness at $\theta_{max}$ & 0.95 \\
    $g_{max}$ / \AA$^{-1}$ & 1.6 \\
    Thickness & Ribbon \\
    Mosaicity & Isotropic incoherent \\
    $R$, $wR$, GoF, $([I > 2\sigma(I)])$ & 0.1049, 0.2501, 1.01 \\
    $R$, $wR$, GoF (all) & 0.1315, 0.2753, 1.14 \\
    No.\ reflections, No.\ of which rejected $(|F_o - F_c| > 4\sigma(F_o))$   & 4241, 12    \\
    No.\ parameters                        & 118                                 \\
    No.\ restraints                        & 0                                  \\
    $\Delta\phi_{\max}$, $\Delta\phi_{\min}$ / $e\,\mathrm{\AA}^{-1}$ & 1.36, $-0.98$ \\
    \bottomrule
\end{tabularx}

  \vspace{0.5em}

\end{table}

\newpage

PXRD was measured on a Bruker D8 Advance diffractometer equipped with a PSD LynxEye detector using Cu–K$\alpha$ radiation ($\lambda = 1.542\,\text{\AA}$). Rietveld refinements and crystalline fraction calculations were performed in Profex 5.5.0.\cite{Doebelin_2015}. The Rietveld refinement as described in the main text (Figure 2) was performed without relaxation of internal coordinates, only the unit cell (a, b, c, $\beta$), isotropic displacement parameter, sample displacement parameter, profile and preferred orientation were refined. The post refinement model gives a = 7.884(2) \text{\AA}, b = 12.615(3) \text{\AA} (fixed), c = 8.460(1) \text{\AA}, $\beta$ = 93.91(7)$^\circ$.

\begin{figure}[htpb]
  \centering
  \includegraphics[width=0.8\textwidth]{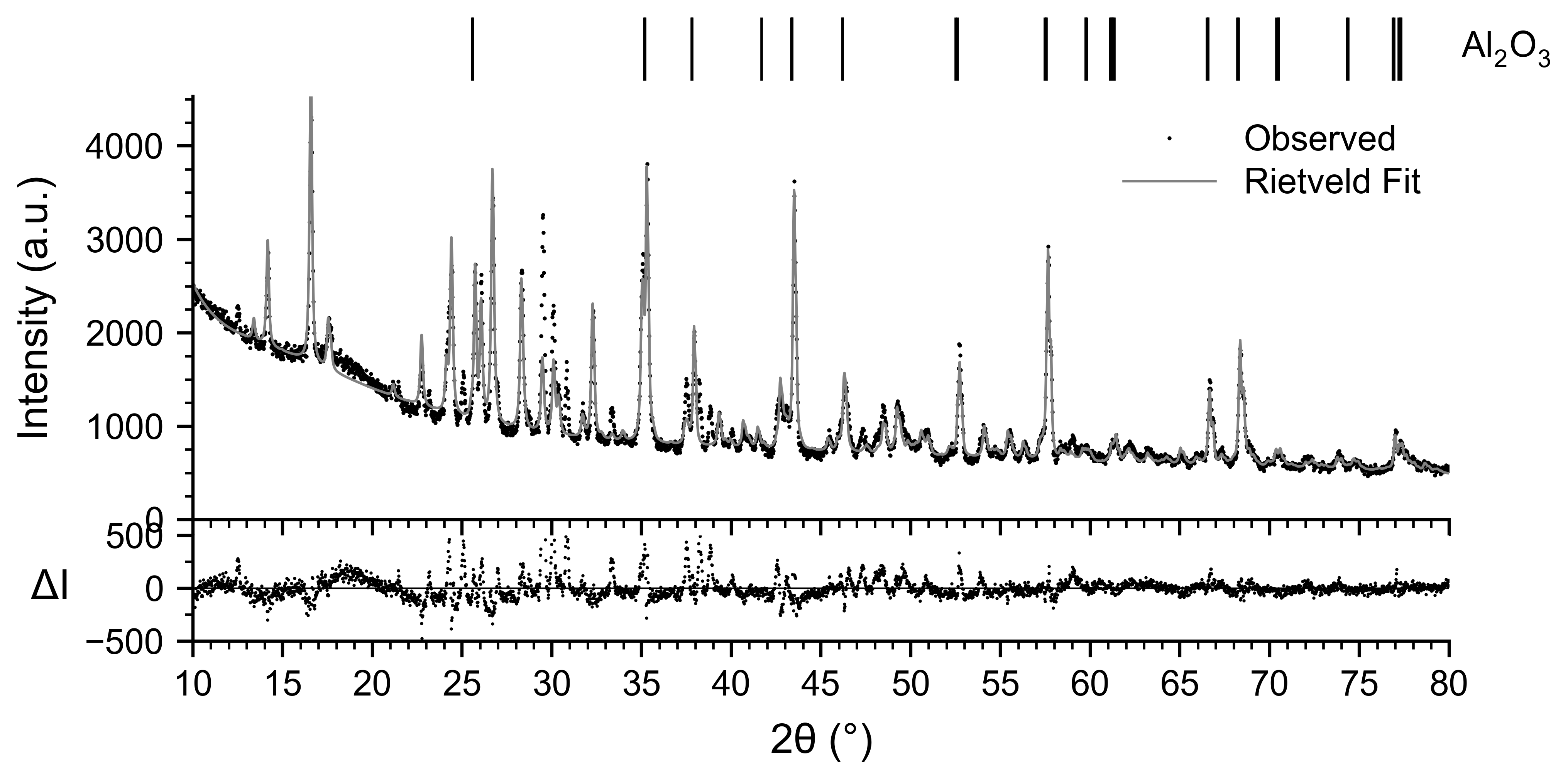}   
  \caption{PXRD pattern of bulk \ruo ~with internal standard $\mathbf{Al_2O_3}$ Internal standard (ICSD 130950). }
  \label{fig:S1_IS}
\end{figure}

For amorphous-phase quantification, $\mathrm{Al_2O_3}$ (3.7 mg, 35.2 wt\%) was mixed with the solid-state \ruo\ product (6.8 mg). The PXRD pattern of this mixture (Figure ~\ref{fig:S1_IS}) was refined using the \ruo and $\mathbf{Al_2O_3}$ (ICSD 130950) structural models, and the refined mass fractions from the Rietveld refinement were converted to the absolute mass fractions in the original sample using the internal standard method in Profex.\cite{Doebelin_2015} This analysis indicates that 27 wt\% of the solid state product is crystalline \ruo.

\section{Magnetic property measurement}

Experiments were carried out on a \emph{Quantum Design MPMS3} magnetic property measurement system
 equipped with the AC and VSM options
(MPMS3 Measurement Release~1.1.16, Build~433; control software
\emph{MultiVu} v2.3.4.26).
A 20\,mg powder sample was wrapped in 4.4\,mg of cling film,
enclosed in gel capsules, and mounted in a straw holder.

\emph{DC mode}  
The sample was cooled to 1.8 K in zero field, a 1 mT field was applied,
and \(M(T)\) was recorded while warming to 300 K at a rate of
1 K\,min\(^{-1}\).
Isothermal \(M(H)\) loops were collected at 1.8 K
(five quadrants, 0 → 7 T → -7 T → 7 T) and at
10, 20, 30, 40, 50, 60, 70, 100, 150, 200, 250, 300 K
(four quadrants, ±7 T).
DC scans used a 30 mm scan length and 4 s scan time.

\emph{VSM mode.}  
The 1 mT \(M(T)\) warming measurement was repeated in VSM operation
with a 5 mm drive amplitude and 2 s averaging time.
Additional \(M(H)\) loops at the same temperature set were measured with
identical VSM settings; for comparison, DC scans were performed with a
10 mm scan length and 1 s scan time.

\section{Scanning and Transmission Electron Microscopy}
\begin{figure}[H]
\centering
\includegraphics[width=0.5\textwidth]{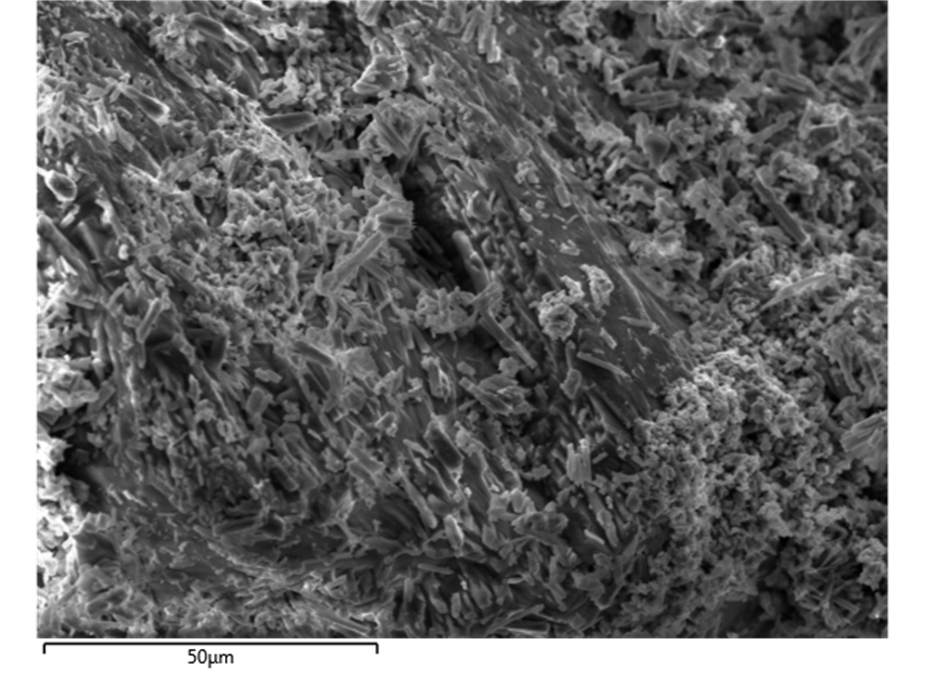}
\caption{Scanning electron micrograph of solid state product of \ruo}
\label{SEM}
\end{figure}

\begin{figure}[H]
\centering
\begin{subfigure}[b]{0.24\textwidth}
\centering
\includegraphics[width=\textwidth]{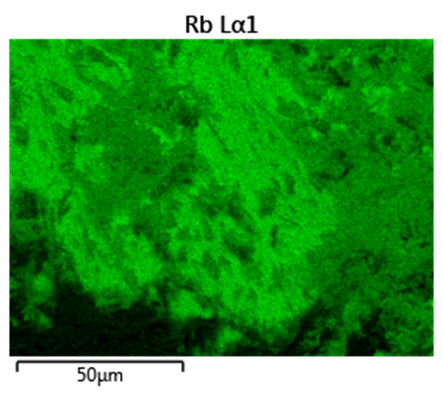}
\caption{Elemental mapping micrograph of Rb}

\end{subfigure}
\hfill
\begin{subfigure}[b]{0.24\textwidth}
\centering
\includegraphics[width=\textwidth]{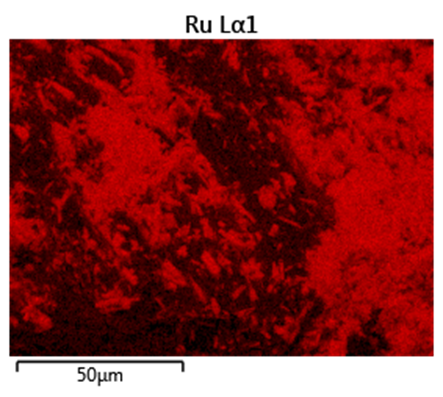}
\caption{Elemental mapping micrograph of Ru}

\end{subfigure}
\hfill
\begin{subfigure}[b]{0.24\textwidth}
\centering
\includegraphics[width=\textwidth]{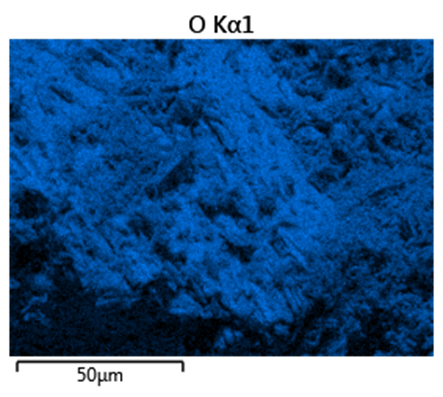}
\caption{Elemental mapping micrograph of O}

\end{subfigure}
\hfill
\begin{subfigure}[b]{0.24\textwidth}
\centering
\includegraphics[width=\textwidth]{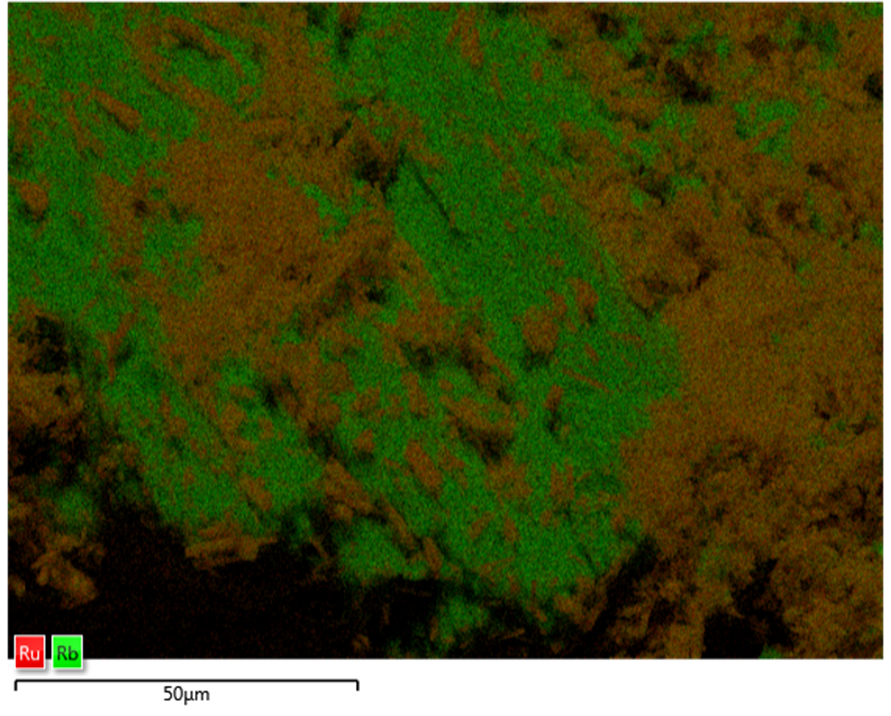}
\caption{Elemental mapping micrograph of O}

\end{subfigure}
\caption{Scanning elemental mapping micrographs of synthesis solid state product}
\label{SEM_EDX}
\end{figure}

SEM analysis was performed on a JEOL JSM-IT300 system. The micrograph as shown in \ref{SEM} revealed two morphologies: larger platelet-like structures, up to 12 $\mu$m in size, mostly located in the rubidium zone, and smaller clusters of particles present in the mixed metal regions. The elemental mapping micrographs in \ref{SEM_EDX} show presence of both Rb and Ru and an overlap of the metals in the solid-state product.

TEM analysis was performed on the JEOL JEM-1400 system reveals rod-like diffracting polycrystalline particles up to $6\,\mu\mathrm{m}$ in length.

\begin{figure}[H]
\centering
\begin{subfigure}[b]{0.4\textwidth}
\centering
\includegraphics[width=\textwidth]{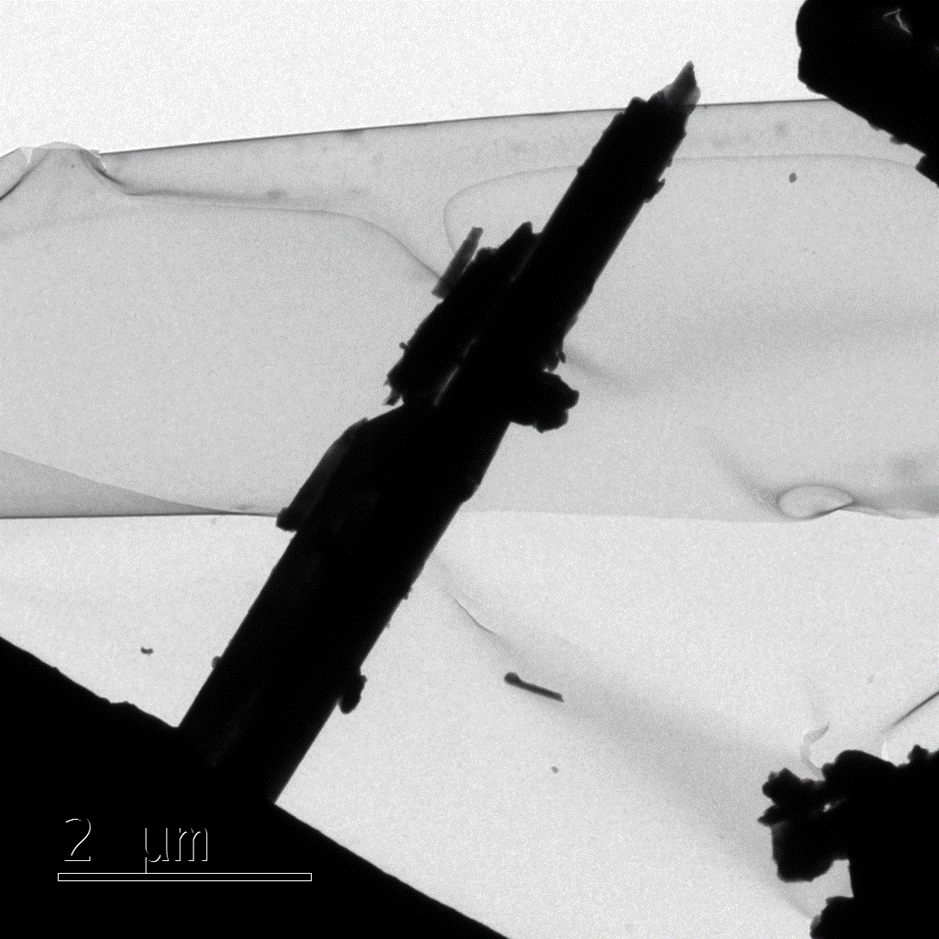}
\caption{Transmission electron micrograph of solid state product crystal.}

\end{subfigure}
\hfill
\begin{subfigure}[b]{0.4\textwidth}
\centering
\includegraphics[width=\textwidth]{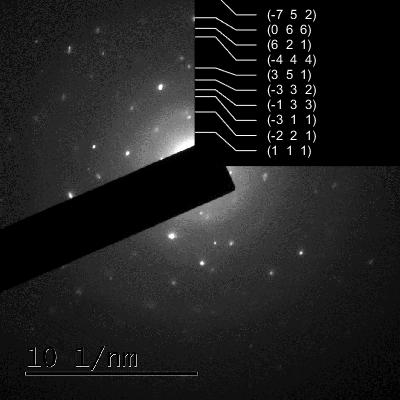}
\caption{Electron diffraction pattern obtained from the solid-state product synthesis.}
\end{subfigure}
\hfill
\caption{TEM analysis of solid-state product}
\label{TEM}
\end{figure}

The rings from the obtained diffraction micrograph were indexed to \ruo (CSD 2514706) using CrysTBox ringGUI, and are listed in Table \ref{tab:ring_id}. \cite{klingercrystbox, klingercrystboxManual} 

\begin{DIFnomarkup}
\begin{table}[htbp]
  \centering
  \caption{Ring identification from the diffraction pattern. Listed are theoretical and measured ring radii and corresponding $d$-spacings.}
  \label{tab:ring_id}
  \begin{tabular}{@{}lcccc@{}}
    \toprule

    \textbf{Plane} &
    \multicolumn{2}{c}{\textbf{Radius / nm$^{-1}$}} &
    \multicolumn{2}{c}{\textbf{$d$-spacing / nm}} \\
    \cmidrule(lr){2-3} \cmidrule(lr){4-5}
    & \textbf{theor.} & \textbf{measured} & \textbf{theor.} & \textbf{measured} \\
    \midrule
    (1 1 1)   & 1.972 & 1.936 & 0.507 & 0.516 \\
    (-2 2 1)  & 3.184 & 3.128 & 0.314 & 0.320 \\
    (-3 1 1)  & 4.024 & 4.022 & 0.249 & 0.249 \\
    (-1 3 3)  & 4.433 & 4.419 & 0.226 & 0.226 \\
    (-3 3 2)  & 5.002 & 4.990 & 0.200 & 0.200 \\
    (3 5 1)   & 5.718 & 5.685 & 0.175 & 0.176 \\
    (-4 4 4)  & 7.492 & 7.497 & 0.133 & 0.133 \\
    (6 2 1)   & 7.994 & 7.993 & 0.125 & 0.125 \\
    (0 6 6)   & 8.623 & 8.614 & 0.116 & 0.116 \\
    (-7 5 2)  & 9.952 & 9.955 & 0.100 & 0.100 \\
    \bottomrule
  \end{tabular}
\end{table}
\end{DIFnomarkup}


\bibliography{bibliography}

@Article{Jeong_2013,
  author    = {Jeong, D. W. and Choi, Hong Chul and Kim, Choong H. and Chang, Seo Hyoung and Sohn, C. H. and Park, H. J. and Kang, T. D. and Cho, Deok-Yong and Baek, S. H. and Eom, C. B. and Shim, J. H. and Yu, J. and Kim, K. W. and Moon, S. J. and Noh, T. W.},
  journal   = {Physical Review Letters},
  title     = {Temperature Evolution of Itinerant Ferromagnetism in SrRuO$_3$ Probed by Optical Spectroscopy},
  year      = {2013},
  issn      = {1079-7114},
  month     = jun,
  number    = {24},
  pages     = {247202},
  volume    = {110},
  doi       = {https://doi.org/10.1103/PhysRevLett.110.247202},
  publisher = {American Physical Society (APS)},
}

@Article{Shepard_1996,
  author    = {Shepard, M. and Cao, G. and McCall, S. and Freibert, F. and Crow, J. E.},
  journal   = {Journal of Applied Physics},
  title     = {Magnetic and transport properties of Na doped SrRuO$_3$ and CaRuO$_3$},
  year      = {1996},
  issn      = {1089-7550},
  month     = apr,
  number    = {8},
  pages     = {4821--4823},
  volume    = {79},
  doi       = {https://doi.org/10.1063/1.361619},
  publisher = {AIP Publishing},
}

@Article{Maeno_1994,
  author    = {Maeno, Y. and Hashimoto, H. and Yoshida, K. and Nishizaki, S. and Fujita, T. and Bednorz, J. G. and Lichtenberg, F.},
  journal   = {Nature},
  title     = {Superconductivity in a layered perovskite without copper},
  year      = {1994},
  issn      = {1476-4687},
  month     = dec,
  number    = {6506},
  pages     = {532--534},
  volume    = {372},
  doi       = {https://doi.org/10.1038/372532a0},
  publisher = {Springer Science and Business Media LLC},
}

@Article{Miura_2007,
  author    = {Miura, Yoko and Yasui, Yukio and Sato, Masatoshi and Igawa, Naoki and Kakurai, Kazuhisa},
  journal   = {Journal of the Physical Society of Japan},
  title     = {New-Type Phase Transition of Li$_2$RuO$_3$ with Honeycomb Structure},
  year      = {2007},
  issn      = {1347-4073},
  month     = mar,
  number    = {3},
  pages     = {033705},
  volume    = {76},
  doi       = {https://doi.org/10.1143/JPSJ.76.033705},
  publisher = {Physical Society of Japan},
}

@Article{Wang_2014,
  author    = {Wang, J. C. and Terzic, J. and Qi, T. F. and Ye, Feng and Yuan, S. J. and Aswartham, S. and Streltsov, S. V. and Khomskii, D. I. and Kaul, R. K. and Cao, G.},
  journal   = {Physical Review B},
  title     = {Lattice-tuned magnetism of Ru$^{4+}$(4d$^4$) ions in single crystals of the layered honeycomb ruthenates Li$_2$RuO$_3$ and Na$_2$RuO$_3$},
  year      = {2014},
  issn      = {1550-235X},
  month     = oct,
  number    = {16},
  pages     = {161110},
  volume    = {90},
  doi       = {https://doi.org/10.1103/PhysRevB.90.161110},
  publisher = {American Physical Society (APS)},
}

@Article{Miura_2009,
  author    = {Miura, Yoko and Sato, Masatoshi and Yamakawa, Youichi and Habaguchi, Tatsuro and Ōno, Yoshiaki},
  journal   = {Journal of the Physical Society of Japan},
  title     = {Structural Transition of Li$_2$RuO$_3$ Induced by Molecular-Orbit Formation},
  year      = {2009},
  issn      = {1347-4073},
  month     = sep,
  number    = {9},
  pages     = {094706},
  volume    = {78},
  doi       = {https://doi.org/10.1143/JPSJ.78.094706},
  publisher = {Physical Society of Japan},
}

@Article{Alexander_2003,
  author    = {Alexander, A. and Battle, P. D. and Burley, J. C. and Gallon, Daniel J. and Grey, Clare P. and Kim, S. H.},
  journal   = {Journal of Materials Chemistry},
  title     = {Structural and magnetic properties of Li$_3$RuO$_4$},
  year      = {2003},
  issn      = {1364-5501},
  number    = {10},
  pages     = {2612},
  volume    = {13},
  doi       = {https://doi.org/10.1039/B305220F},
  publisher = {Royal Society of Chemistry (RSC)},
}

@Article{Doebelin_2015,
  author    = {Doebelin, Nicola and Kleeberg, Reinhard},
  journal   = {Journal of Applied Crystallography},
  title     = {Profex: a graphical user interface for the Rietveld refinement program BGMN},
  year      = {2015},
  issn      = {1600-5767},
  month     = aug,
  number    = {5},
  pages     = {1573--1580},
  volume    = {48},
  doi       = {https://doi.org/10.1107/S1600576715014685},
  publisher = {International Union of Crystallography (IUCr)},
}

@Article{3DED_1,
  author = {},
  title  = {Rigaku Oxford Diffraction, CrysAlis PRO},
  year   = {2025},
}

@Article{Sheldrick_2015,
  author    = {Sheldrick, George M.},
  journal   = {Acta Crystallographica Section A Foundations and Advances},
  title     = {SHELXT– Integrated space-group and crystal-structure determination},
  year      = {2015},
  issn      = {2053-2733},
  month     = jan,
  number    = {1},
  pages     = {3--8},
  volume    = {71},
  doi       = {10.1107/S2053273314026370},
  publisher = {International Union of Crystallography (IUCr)},
}

@Article{Bourhis_2015,
  author    = {Bourhis, Luc J. and Dolomanov, Oleg V. and Gildea, Richard J. and Howard, Judith A. K. and Puschmann, Horst},
  journal   = {Acta Crystallographica Section A Foundations and Advances},
  title     = {The anatomy of a comprehensive constrained, restrained refinement program for the modern computing environment –Olex2 dissected},
  year      = {2015},
  issn      = {2053-2733},
  month     = jan,
  number    = {1},
  pages     = {59--75},
  volume    = {71},
  doi       = {doi:10.1107/S2053273314022207},
  publisher = {International Union of Crystallography (IUCr)},
}

@Article{Dolomanov_2009,
  author    = {Dolomanov, Oleg V. and Bourhis, Luc J. and Gildea, Richard J. and Howard, Judith A. K. and Puschmann, Horst},
  journal   = {Journal of Applied Crystallography},
  title     = {OLEX2: a complete structure solution, refinement and analysis program},
  year      = {2009},
  issn      = {0021-8898},
  month     = jan,
  number    = {2},
  pages     = {339--341},
  volume    = {42},
  doi       = {http://dx.doi.org/10.1107/S0021889808042726},
  publisher = {International Union of Crystallography (IUCr)},
}

@Article{Saha_2022,
  author    = {Saha, Ambarneil and Nia, Shervin S. and Rodríguez, José A.},
  journal   = {Chemical Reviews},
  title     = {Electron Diffraction of 3D Molecular Crystals},
  year      = {2022},
  issn      = {1520-6890},
  month     = aug,
  number    = {17},
  pages     = {13883--13914},
  volume    = {122},
  doi       = {https://doi.org/10.1021/acs.chemrev.1c00879},
  publisher = {American Chemical Society (ACS)},
}

@Article{Brese_1991,
  author    = {Brese, N. E. and O’Keeffe, M.},
  journal   = {Acta Crystallographica Section B Structural Science},
  title     = {Bond-valence parameters for solids},
  year      = {1991},
  issn      = {0108-7681},
  month     = apr,
  number    = {2},
  pages     = {192--197},
  volume    = {47},
  doi       = {https://doi.org/10.1107/S0108768190011041},
  publisher = {International Union of Crystallography (IUCr)},
}

@Article{Gagn__2015,
  author    = {Gagné, Olivier Charles and Hawthorne, Frank Christopher},
  journal   = {Acta Crystallographica Section B Structural Science, Crystal Engineering and Materials},
  title     = {Comprehensive derivation of bond-valence parameters for ion pairs involving oxygen},
  year      = {2015},
  issn      = {2052-5206},
  month     = sep,
  number    = {5},
  pages     = {562--578},
  volume    = {71},
  doi       = {https://doi.org/10.1107/S2052520615016297},
  publisher = {International Union of Crystallography (IUCr)},
}

@Article{Gagn__2020,
  author    = {Gagné, Olivier Charles and Hawthorne, Frank Christopher},
  journal   = {IUCrJ},
  title     = {Bond-length distributions for ions bonded to oxygen: results for the transition metals and quantification of the factors underlying bond-length variation in inorganic solids},
  year      = {2020},
  issn      = {2052-2525},
  month     = jun,
  number    = {4},
  pages     = {581--629},
  volume    = {7},
  doi       = {https://doi.org/10.1107/S2052252520005928},
  publisher = {International Union of Crystallography (IUCr)},
}

@Article{Reeves_2019,
  author    = {Reeves, Philip J. and Seymour, Ieuan D. and Griffith, Kent J. and Grey, Clare P.},
  journal   = {Chemistry of Materials},
  title     = {Characterizing the Structure and Phase Transition of Li$_2$RuO$_3$ Using Variable-Temperature $^{17}$O and $^{7}$Li NMR Spectroscopy},
  year      = {2019},
  issn      = {1520-5002},
  month     = mar,
  number    = {8},
  pages     = {2814--2821},
  volume    = {31},
  doi       = {https://doi.org/10.1021/acs.chemmater.8b05178},
  publisher = {American Chemical Society (ACS)},
}

@Article{Sanjeewa_2020,
  author    = {Sanjeewa, Liurukara D. and Liu, Yaohua and Xing, Jie and Fishman, Randy S. and Kolambage, Mudithangani T. K. and McGuire, Michael A. and McMillen, Colin D. and Kolis, Joseph W. and Sefat, Athena S.},
  journal   = {physica status solidi (b)},
  title     = {Stacking Faults and Short‐Range Magnetic Correlations in Single Crystal Y$_5$Ru$_2$O$_{12}$: A Structure with Ru$^{+4.5}$ One‐Dimensional Chains},
  year      = {2020},
  issn      = {1521-3951},
  month     = sep,
  number    = {2},
  volume    = {258},
  doi       = {10.1002/pssb.202000197},
  publisher = {Wiley},
}

@Article{Pet_ek_2023,
  author    = {Petříček, Václav and Palatinus, Lukáš and Plášil, Jakub and Dušek, Michal},
  journal   = {Zeitschrift für Kristallographie - Crystalline Materials},
  title     = {Jana2020 – a new version of the crystallographic computing system Jana},
  year      = {2023},
  issn      = {2196-7105},
  month     = apr,
  number    = {7–8},
  pages     = {271--282},
  volume    = {238},
  doi       = {https://doi.org/10.1515/zkri-2023-0005
},
  publisher = {Walter de Gruyter GmbH},
}

@article{klingercrystbox,
  author = "Klinger, M. and Jäger, A.",
  title = "{Crystallographic Tool Box (CrysTBox): automated tools for transmission electron microscopists and crystallographers}",
  journal = "Journal of Applied Crystallography",
  year = "2015",
  volume = "48",
  number = "6",
  pages = "",
  month = "Dec",
  doi = {10.1107/S1600576715017252},
  url = {http://dx.doi.org/10.1107/S1600576715017252},
}

@book{klingercrystboxManual,
  title={CrysTBox - Crystallographic Toolbox},
  author={Klinger, M.},
  isbn={978-80-905962-3-8},
  url={http://www.fzu.cz/~klinger/crystbox.pdf},
  year={2015},
  address = {Prague},
  publisher={Institute of Physics of the Czech Academy of Sciences},
}
\end{document}